\documentclass[final]{svjour3}
\usepackage{rotating}
\usepackage{amssymb}
\usepackage{mathptmx}
\usepackage{hyperref}
\hypersetup{
     colorlinks   = true,
     allcolors    =blue
}
\usepackage{graphicx}
\usepackage{caption}
\usepackage{tabularx}
\usepackage[numbers, square, sort]{natbib}

\usepackage{soul}

\usepackage{xcolor}

\makeatletter
\journalname{Journal of Low Temperature Physics}
\usepackage{etoolbox}

\begin{document}


\newcommand{\hdblarrow}{H\makebox[0.9ex][l]{$\downdownarrows$}-}
\title{Sensitivity of the Prime-Cam Instrument on the CCAT-prime Telescope}

\author{S.K. Choi \textsuperscript{1} \and 
J. Austermann \textsuperscript{2}  \and  
K. Basu \textsuperscript{3} \and
N. Battaglia \textsuperscript{1} \and
F. Bertoldi \textsuperscript{3} \and
D.T. Chung \textsuperscript{4} \and   
N.F. Cothard \textsuperscript{5} \and 
S. Duff \textsuperscript{2}  \and  
C.J. Duell \textsuperscript{5} \and 
P.A. Gallardo \textsuperscript{5} \and 
J. Gao \textsuperscript{2}  \and  
T. Herter \textsuperscript{1} \and 
J. Hubmayr \textsuperscript{2} \and 
M.D. Niemack \textsuperscript{5} \and 
T. Nikola \textsuperscript{6} \and 
D. Riechers \textsuperscript{1} \and 
K. Rossi \textsuperscript{6} \and 
G.J. Stacey \textsuperscript{1} \and 
J.R. Stevens \textsuperscript{5} \and 
E.M. Vavagiakis \textsuperscript{5} \and
M. Vissers \textsuperscript{2} \and
S. Walker \textsuperscript{2,7}
}

\institute{\textsuperscript{1}Department of Astronomy, Cornell University, Ithaca, NY 14853, USA \\
\textsuperscript{2}Quantum Sensors Group, NIST, Boulder, CO 80305, USA \\
\textsuperscript{3}Argelander Institute for Astronomy, University of Bonn, D-53121 Bonn, Germany \\
\textsuperscript{4}Kavli Institute for Particle Astrophysics and Cosmology \& Physics Department, Stanford University, Stanford, CA 94305, USA \\
\textsuperscript{5}Department of Physics, Cornell University, Ithaca, NY 14853, USA \\
\textsuperscript{6}Cornell Center for Astrophysics and Planetary Science, Cornell University, Ithaca, NY 14853, USA \\
\textsuperscript{7}Center for Astrophysics and Space Astronomy, University of Colorado Boulder, Boulder, CO 80309, USA\\
\email{skc98@cornell.edu}
}

\authorrunning
\titlerunning
\maketitle

\begin{abstract}

CCAT-prime is a new 6 m crossed Dragone telescope designed to characterize the Cosmic Microwave Background (CMB) polarization and foregrounds, measure the Sunyaev-Zel'dovich effects of galaxy clusters, map the [CII] emission intensity from the Epoch of Reionization (EoR), and monitor accretion luminosity over multi-year timescales of hundreds of protostars in the Milky Way. CCAT-prime will make observations from a 5,600 m altitude site on Cerro Chajnantor in the Atacama Desert of northern Chile. The novel optical design of the telescope combined with high surface accuracy ($<$10 $\mu$m) mirrors and the exceptional atmospheric conditions of the site will enable sensitive broadband, polarimetric, and spectroscopic surveys at sub-mm to mm wavelengths. Prime-Cam, the first light instrument for CCAT-prime, consists of a 1.8 m diameter cryostat that can house seven individual instrument modules. Each instrument module, optimized for a specific science goal, will use state-of-the-art kinetic inductance detector (KID) arrays operated at $\sim$100 mK, and Fabry-Perot interferometers (FPI) for the EoR science. Prime-Cam will be commissioned with staged deployments to populate the seven instrument modules. The full instrument will consist of 60,000 polarimetric KIDs at a combination of 220/280/350/410 GHz, 31,000 KIDS at 250/360 GHz coupled with FPIs, and 21,000 polarimetric KIDs at 850 GHz. Prime-Cam is currently being built, and the CCAT-prime telescope is designed and under construction by Vertex Antennentechnik GmbH to achieve first light in 2021. CCAT-prime is also a potential telescope platform for the future CMB Stage-IV observations.

\end{abstract}

\section{Introduction}
\label{sec:intro}
CCAT-prime is a new 6 m crossed Dragone telescope currently under construction to be located on Cerro Chajnantor in the Atacama Desert for first light in 2021. CCAT-prime is designed to tackle many important questions in modern cosmology and astrophysics, from constraining extensions to the $\Lambda$CDM model to the physics of star formation. These questions are addressed with sensitive broadband, polarimetric, and spectroscopic observations at mm to sub-mm wavelengths from an exceptional 5,600 m altitude observing site in the Atacama Desert with high atmospheric transmission. At $>$300 GHz, CCAT-prime will have $>$2 times faster mapping speed than ongoing and currently planned experiments \cite{stacey/2018, aravena/2019}. A novel optical design of the telescope provides a large flat focal plane that will be filled with state-of-the-art low temperature detectors. These detectors have background-limited sensitivity and are scaled to large formats to increase detector count and focal plane filling fraction.

The key science questions Prime-Cam will help address include: 1) what is the fundamental physics of the early universe? 2) what is the nature of dark energy and the sum of the neutrino masses? 3) what are the properties of the sources of reionization? 4) how does star formation proceed in the Milky Way and in nearby galaxies? Prime-Cam will make the most sensitive broadband measurements to date at $>$200 GHz to detect Rayleigh scattering of the CMB \cite{lewis/2013, alipour/2015} to improve constraints on cosmological parameters including the number of relativistic species, optical depth, and neutrino mass by breaking degeneracies that exist when using primary CMB constraints only. Measurements from Prime-Cam combined with those from the Atacama Cosmology Telescope \cite{henderson/2016b} (ACT) and Simons Observatory \cite{SO/2019} (SO) will detect $\sim$16,000 galaxy clusters via the Sunyaev-Zel'dovich (SZ) effect. Peculiar motions of these clusters will be sensitive to dark energy \cite{mueller/2015a} and the sum of the neutrino masses \cite{mueller/2015b}. Prime-Cam's high frequency measurements will enable temperature measurements in individual clusters from the relativistic SZ effect \cite{erler/2018, mittal/2018}, and also improve the constraints of primordial B-mode measurements by better characterizing the galactic dust polarization. High cadence wide-field continuum imaging will also permit searches for transients that include rapid changes in accretion rates in nearby protostars. Prime-Cam's spectroscopic survey in the 210--420 GHz band will map the line intensity of [CII] to improve our understanding of the first star formation and galaxy evolution throughout the epoch of reionization \cite{gong/2012}. [CII] line intensity mapping will provide a measurement of the three-dimensional high redshift universe to also search for signature of dark matter from decays and annihilation \cite{Creque-Sarbinowski/2018} and potentially probe primordial non-Gaussianity of different models within early universe theories \cite{komatsu/2009, Moradinezhad/2019, kovetz/2019}.


Advances in contemporary low temperature detectors for mm to sub-mm wavelength observations continue to push the state-of-the-art sensitivities. Transition edge sensors (TES) have reached background-limited sensitivity, and increased detector count and frequency channels using multichroic arrays on 150 mm silicon wafers are the state-of-the-art. TES arrays are currently read out using time domain or frequency domain multiplexing (mux) with a mux factor of 64 \cite{henderson/2016a, bender/2016}; the current goal is to increase this mux factor considerably \cite{galitzki/2018}. Kinetic inductance detectors (KIDs) are naturally multiplexed to a factor of $>$500 \cite{austermann/2018, lourie/2018}, which is currently limited by the warm readout electronics. Given the relatively simpler readout architecture and the scalability, KIDs are emerging to be a mainstream choice of sensors in mm to sub-mm wavelength observations \cite{masi/2019, catalano/2018}.

Some of the ongoing or currently planned large aperture telescope experiments at mm to sub-mm include ACT \cite{henderson/2016b}, the South Pole Telescope \cite{carlstrom/2011}, the SO large aperture telescope \cite{SO/2019}, TIME \cite{crites/2014}, CONCERTO \cite{lagache/2018}, and OLIMPO \cite{masi/2019}. With the supreme observing site for CCAT-prime combined with high surface-accuracy mirrors of the telescope and a large focal plane, Prime-Cam will have unparalleled sensitivity and dynamic range of angular scales above 200 GHz.


\begin{figure}[t]
    \centering
    \includegraphics[width=0.8\textwidth]{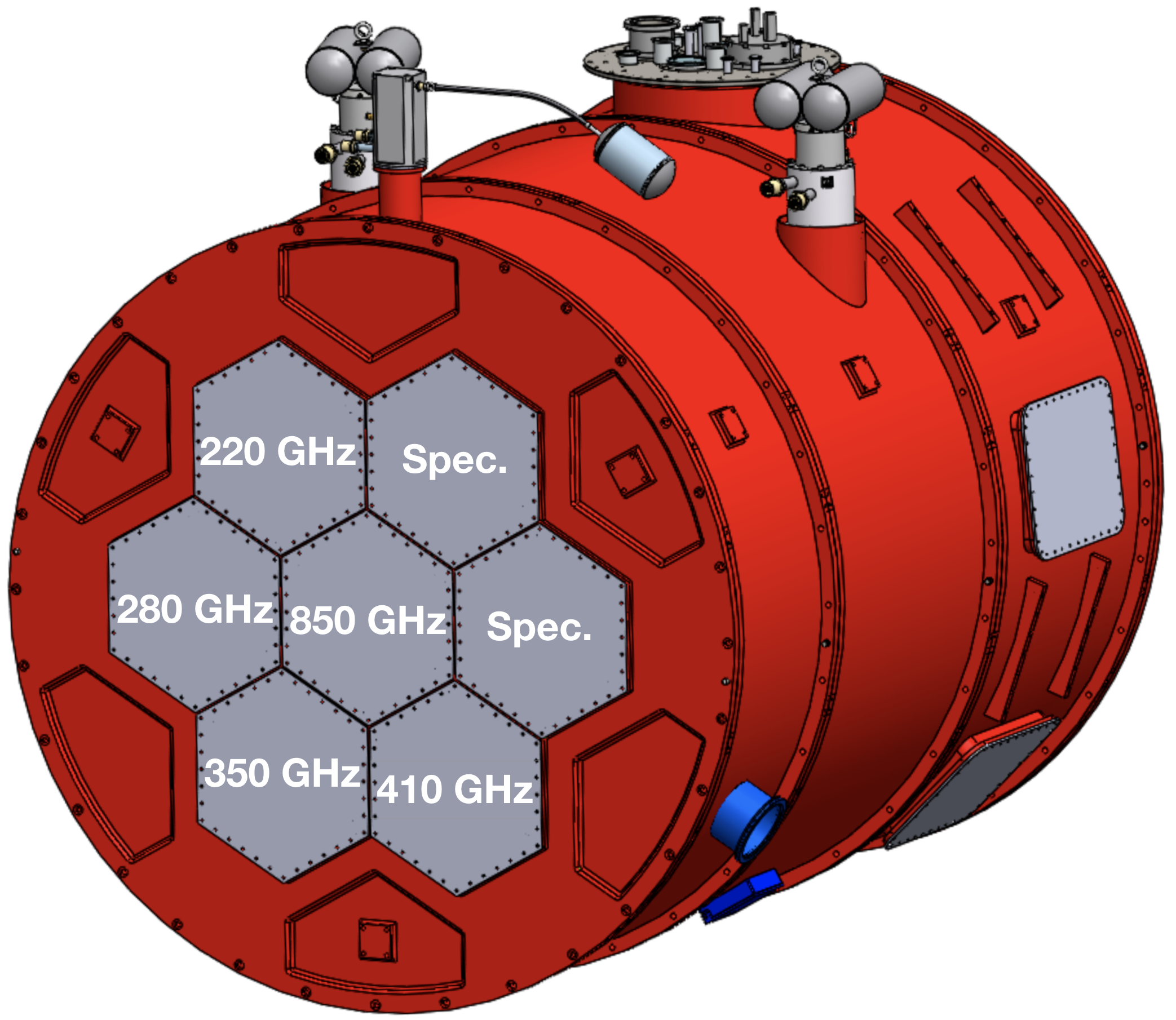}
    \caption{A model of the Prime-Cam cryostat (1.8 m diameter) is shown with a possible configuration of broadband and spectrometer instrument module positions.}
\label{fig:prime_cam}
\end{figure}

\section{Instrument}
\label{sec:inst}
CCAT-prime is an off-axis crossed Dragone telescope with a 6 m primary and a 5.7 m secondary mirror \cite{parshley/2018b}. CCAT-prime's novel optics achieve a wide diffraction-limited field of view (from 8 deg at 100 GHz to 2 deg at 860 GHz) and large flat focal plane area, enabling illumination of about ten times more detectors than any current mm telescope \cite{niemack/2016}. CCAT-prime's off-axis design also achieves low emissivity ($<$2$\%$), which reduces loading to detectors and improves sensitivity \cite{parshley/2018a}. 

Prime-Cam is a first light receiver in development for CCAT-prime \cite{vavagiakis/2018}. Prime-Cam, shown in Figure~\ref{fig:prime_cam}, is a 1.8 m diameter cryostat housing seven independent instrument modules (or optics tubes). Each instrument module is optimized for a particular observing/science band. Prime-Cam will have broadband instrument modules sensitive to five different frequencies (220, 280, 350, 410, 850 GHz) and spectroscopic instrument modules sensitive to 210--420 GHz with $R=100$. The optics tube designs are evolving from collaborative work with SO \cite{dicker/2018, vavagiakis/2018}, and will contain metal-mesh infrared blocking filters, absorbing filters, and meta-material anti-reflection coated silicon lenses \cite{datta/2013} designed for all observing bands. Each instrument module will be populated with three detector arrays fabricated on 150-mm silicon wafers. Prime-Cam will be cooled with three pulse tube coolers and a dilution refrigerator to operate the detectors at $\sim$100 mK. 

\begin{figure}
    \centering
    \includegraphics[width=1\textwidth]{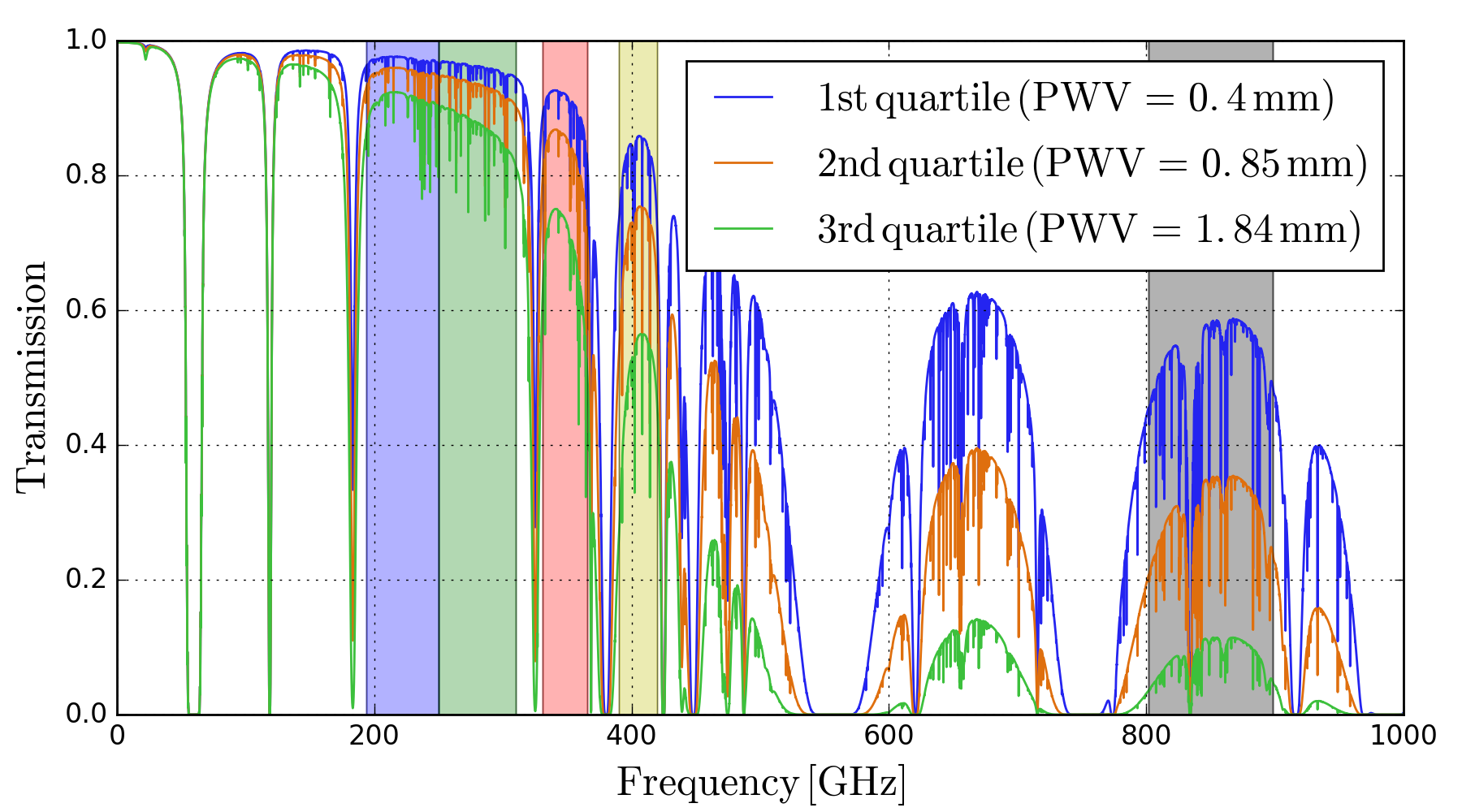}
    \caption{Atmospheric transmission spectra calculated with the \texttt{am} code are shown in solid lines for three quartiles of PWV values measured in the CCAT-prime site \cite{radford/2016}. The PWV values in the caption take into account an observation elevation of 45 deg. Colored bands show the approximate frequency bands of the broadband modules in design. Spectrometer will cover 210--420 GHz.}
\label{fig:atm_trans}
\end{figure}

Current mm observations draw strong heritage from TES arrays read out with SQUID multiplexing. Background-limited sensitivity and large format arrays with multi-chroic observing bands make TES arrays an attractive choice for mm observations given the optimal detector spacings. However, at higher frequencies, better sensitivities can be achieved with higher detector count, and naturally-multiplexed KID arrays are already pushing the pixel densities beyond those of the densest TES arrays currently in use \cite{austermann/2018, duff/2016, ho/2016}. Given the simpler readout architecture and integration procedure required, Prime-Cam is baselining KID arrays for the broadband measurements based on designs from TolTEC KID arrays \cite{austermann/2018}. The current best warm readout systems (ROACH-2) multiplex KIDs at a factor of $\sim$600, but the next generation RFSoC\footnote{Xilinx RFSoC website: https://www.xilinx.com/products/silicon-devices/soc/rfsoc.html} systems capable of reading out $\sim$5,000 KIDs are in development. 

Spectroscopic modules will use cold Fabry-Perot Interferometers (FPI) installed at the 4 K Lyot stop. Broadband KID arrays with frequency centers at 250 GHz and 360 GHz sensitive to the 2nd and 3rd order of the fringes are currently being designed. More details are found in \cite{cothard/2019}.

A fully populated configuration for Prime-Cam under study includes broadband polarimetric instrument modules at 220, 280, 350, 410, and 850 GHz, and two spectroscopic modules sensitive to 210--420 GHz. The current design for the 280 GHz KID array allows tiling of three 150 mm KID arrays, each with $\sim$3500 sensors in a single instrument module. Scaling from this, assuming a mux factor of  $\leq$580 per microwave feed line, Prime-Cam will have $\sim$8,000 KIDs at 220 GHz, $\sim$10,000 KIDs at 280 GHz, and $\sim$21,000 KIDs each at 350, 410, and 850 GHz. Spectroscopic modules will have $\sim$10,000 KIDs at 250 GHz and $\sim$21,000 KIDs at 360 GHz. All detector spacings in design here are at $\geq$1.1$F\lambda$. An optimization study of different frequency bands to maximize science return is ongoing \cite{battaglia/2019} and may evolve this current baseline Prime-Cam configuration. 

\begin{table}
\begin{center}
\setlength{\tabcolsep}{8pt}
Broadband channels wide survey (15,000 deg$^2$; 4,000 hours)
\begin{tabularx}{1\textwidth}{c c c c c c c c}
$\nu$ & $\Delta\nu$ & Resolution & NEI &Sensitivity & NET & $N_\mathrm{white}$ & $N_\mathrm{red}$ \\
GHz & GHz & arcsec  & Jy sr$^{-1}\sqrt{s}$ & $\mu$K-arcmin  & $\mu$K$\sqrt{s}$ & $\mu$K$^2$ & $\mu$K$^2$ \\
 \hline
 \hline
220 & 56 & 57 & 3,700 & 15  & 7.6 & 1.8$\times 10^{-5}$ & 1.6$\times 10^{-2}$\\
280 & 60 & 45 & 6,100 & 27  & 14 & 6.4$\times 10^{-5}$ & 1.1$\times 10^{-1}$\\
350 & 35 & 35 & 16,500 & 105   & 54 & 9.3$\times 10^{-4}$ & 2.7$\times 10^{0}$\\
410 & 30 & 30 & 39,400 & 372  & 192 & 1.2$\times 10^{-2}$ & 1.7$\times 10^{1}$\\
850 & 97 & 14 & 6.0$\times 10^{7}$\textsuperscript{$\dagger$} & 5.7$\times 10^5$ & 3.0$\times 10^5$ & 2.8$\times 10^{4}$ & 6.1$\times 10^{6}$\\ 
 \hline
\vspace{0.05in}
 \end{tabularx}
Broadband channels star formation survey in 1st quartile PWV (410 deg$^2$; 680 hours)
 \begin{tabularx}{1\textwidth}{c c c c c c c c}
$\nu$ & $\Delta\nu$ & Resolution & NEI &Sensitivity & NET  & $N_\mathrm{white}$ & $N_\mathrm{red}$ \\
GHz & GHz & arcsec  & Jy sr$^{-1}\sqrt{s}$ & $\mu$K-arcmin  & $\mu$K$\sqrt{s}$ & $\mu$K$^2$ & $\mu$K$^2$ \\
 \hline
 \hline
220 & 56 & 57 & 3,000 & 6  & 6.3 & 2.9$\times 10^{-6}$ & 2.5$\times 10^{-3}$\\
280 & 60 & 45 & 4,900 & 11  & 11 & 1.0$\times 10^{-5}$ & 1.7$\times 10^{-2}$\\
350 & 35 & 35 & 12,300 & 42   & 40 & 1.5$\times 10^{-4}$ & 4.3$\times 10^{-1}$\\
410 & 30 & 30 & 27,400 & 149  & 134 & 1.9$\times 10^{-3}$ & 2.7$\times 10^{0}$\\
850 & 97 & 14 & 3.8$\times 10^{7}$\textsuperscript{$\dagger$} & 2.3$\times 10^5$   & 1.9$\times 10^5$ & 4.5$\times 10^{3}$ & 9.8$\times 10^{5}$\\ 
\hline
\vspace{0.05in}
\end{tabularx}
\setlength{\tabcolsep}{14pt}
Selected spectrometer channels targeted survey (8 deg$^2$; 4,000 hours)
\begin{tabularx}{1\textwidth}{c c c c c c c c c}
$\nu$ & $\Delta\nu$\textsuperscript{*} & Resolution & [CII] redshift & NEI & $N_\mathrm{white}$\\
GHz & GHz & arcsec &  & Jy sr$^{-1}\sqrt{s}$ & Mpc$^3$ Jy$^2$ sr$^{-2}$  \\
 \hline
 \hline
220 & 2.2 & 57 & 7.5  & 12,900 & 1.2$\times 10^{9}$ \\
280 & 2.8 & 45 & 5.8  & 16,600 & 2.0$\times 10^{9}$ \\
350 & 3.5 & 35 & 4.4  & 30,600 & 6.3$\times 10^{9}$ \\
410 & 4.1 & 30 & 3.7  & 61,500 & 2.3$\times 10^{10}$ \\
 \hline
\end{tabularx}
\caption{Top table gives the properties, instantaneous sensitivities, and noise power spectrum parameters of broadband channels for wide area survey (15,000 deg$^2$). All temperatures are given in CMB temperature units. These sensitivities are representative of the first three quartile weighted average. Middle table gives the equivalent information as the top table but for a small field survey planned (410 deg$^2$) for observation days with 1st quartile PWV values. Bottom table gives properties and sensitivities of the four selected frequencies for the spectrometer instrument for a targeted survey (8 deg$^2$), again as the first three quartile weighted average.\\ 
{\small \textsuperscript{$\dagger$}For 850 GHz, instantaneous sensitivity NEI is given for each detector; integrated noise levels in the next columns are given for $\sim$16,600 yielded detectors.\\
\textsuperscript{*}The resolution bandwidth is given by $\Delta\nu = \nu/R$; the noise bandwidth is $\pi/2$ larger since the spectral profile is Lorentzian in form for an FPI.}
}
\label{tab:sens}
\end{center}
\end{table}

\section{Prime-Cam Sensitivity}
\label{sec:sens}

We estimate the expected Prime-Cam sensitivity and calculate the angular noise power spectra. We first estimate the noise equivalent power (NEP) per detector due to photon shot noise and bunching (wave noise), given the instrument emissivity \cite{stacey/2018, parshley/2018a}, sky transmission (Figure~\ref{fig:atm_trans}), and efficiencies (cold transmission: 86--56\%; warm emissivity for 1st quartile weather: 0.08--0.52; detector quantum efficiency: 80\%; aperture efficiency: 72\%). Equations are given in \cite{stacey/2014}. A few notable parameters that distinguish the sensitivity for Prime-Cam are low telescope emissivity ($<$2$\%$), high atmospheric transmission, and large detector count. Atmospheric transmission is calculated from the \texttt{am} code \cite{am/2018} for typical precipitable water vapor (PWV) values measured in the CCAT-prime site \cite{radford/2016}. Considering a typical observation elevation of 45 deg, atmospheric transmission for the first three quartiles of PWV values are shown along with the Prime-Cam observing bands in Figure~\ref{fig:atm_trans}.

We assume photon noise limited sensitivity and compute the detector array noise equivalent temperature (NET) from the NEPs and the detector counts given in Section~\ref{sec:inst}. Given that our detector spacings are $\geq$1.1$F\lambda$, we expect the white noise degradation due to optical correlation is insignificant at these high frequencies \cite{hill/2018}. With the expected observation time and survey areas for different instrument modules, we calculate the integrated map depth, and corresponding angular noise power spectrum at each frequency given the beam size. 

\begin{figure}
    \centering
    \includegraphics[width=1\textwidth]{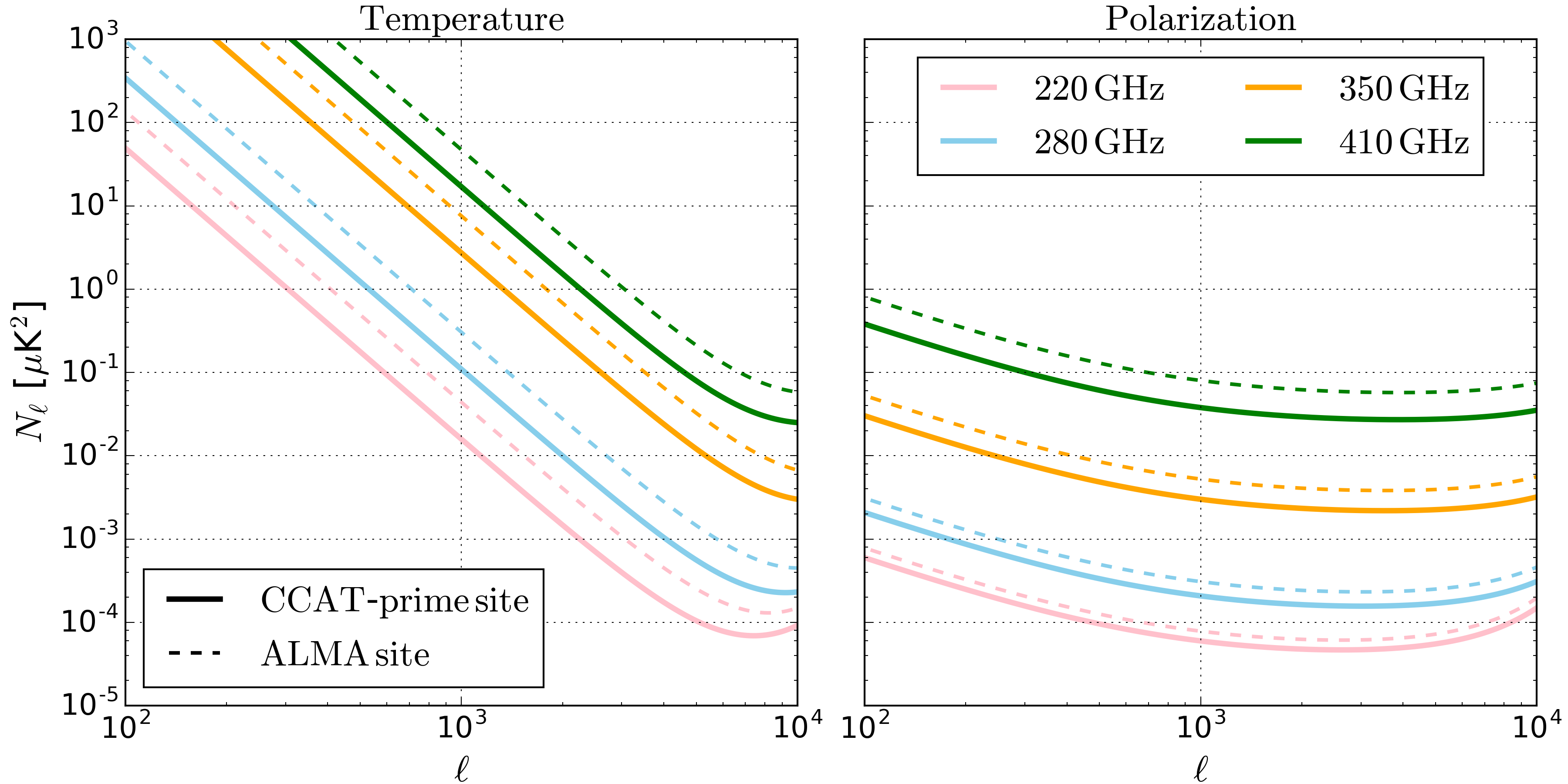}
    \caption{Angular noise power spectra for temperature (left) and polarization (right) are shown. Different colors indicate the four broadband frequency channels. Solid lines indicate the expected noise power spectra for the CCAT-prime site. Dashed lines show the representative noise power spectra at the ALMA site, calculated with higher PWVs based on measurements \cite{radford/2016}.}
\label{fig:noise_curve}
\end{figure}

Large scale ($1/f$) noise in temperature maps in ground based mm observations is dominated by the fluctuation of the atmosphere \cite{louis/2017}. The $1/f$ or $1/\ell$ noise for Prime-Cam frequency bands are estimated using the observed ACT $1/\ell$ noise at 150 GHz extrapolated to higher frequencies using the SO noise model \cite{SO/2019}:
$$N_\ell = N_\mathrm{red} \bigg(\frac{\ell}{\ell_\mathrm{knee}}\bigg)^{\alpha_\mathrm{knee}} + N_\mathrm{white},$$
where $N_\ell$ is the total angular noise power spectrum, where $N_\mathrm{white}$ is the white noise component, and $N_\mathrm{red}$, $\ell_\mathrm{knee}=1000$, and $\alpha_\mathrm{knee}=-3.5$ parametrize the large scale atmospheric noise component. Extrapolation of the ACT $N_\mathrm{red}$ at 150 GHz to Prime-Cam frequencies are done with factors calculated from the ratio of the PWV dependent Rayleigh-Jeans (RJ) temperature spectrum from the \texttt{am} code at the Prime-Cam frequency to 150 GHz, following the SO noise model \cite{SO/2019}. The polarization noise power spectrum is given by $N_\mathrm{red} = N_\mathrm{white}$ and $\ell_\mathrm{knee} = 700$ and $\alpha_\mathrm{knee} = -1.4$ following the SO model, which is based on ACT observations. The resulting temperature and polarization noise power spectra are shown in Figure~\ref{fig:noise_curve}. We also show in Figure~\ref{fig:noise_curve} the representative noise power spectra for the ALMA site, calculated with higher PWVs based on measurements \cite{radford/2016}. For spectroscopic observations, we give the instantaneous sensitivity and the white noise level. With strong spectral correlations, atmospheric noise can possibly be mitigated with component separation using many spectrometer channels. We will investigate the importance of $N_\mathrm{red}$ for spectrometer channels in the near future, building on earlier studies of $1/f$ noise for HI intensity mapping experiments \cite{harper/2018, chen/2019}. Prime-Cam sensitivity and parameters for noise power spectra are summarized in Table~\ref{tab:sens}.


%



\section{Conclusion}
We presented the expected sensitivity of the Prime-Cam receiver on the CCAT-prime telescope. With the state-of-the-art telescope at an outstanding 5,600 m altitude observing site, Prime-Cam will have unprecedented sensitivity for broadband and spectroscopic observations at $>$220 GHz. Detailed forecasts of CCAT-prime science goals will be presented based on the expected sensitivity of Prime-Cam described in this proceeding \cite{battaglia/2019}.

\begin{acknowledgements}
SKC acknowledges support from the Cornell Presidential Postdoctoral Fellowship. MDN acknowledges support from NSF award AST-1454881. Work by NFC was supported by NASA Space Technology Research Fellowship.
\end{acknowledgements}

\bibliographystyle{unsrt85}
\bibliography{choi_LTD18_200224.bbl}

\end{document}